# Long term performance of the SwissQuantum quantum key distribution network in a field environment

**D Stucki[1], M Legré[2]‡, F Buntschu[3], B Clausen[2], N Felber[4], N Gisin[1], L Henzen[4], P Junod[6], G Litzistorf[5], P Monbaron[6], L Monat[2], J-B Page[2], D Perroud[3], G Ribordy[2], A Rochas[2], S Robyr[2], J Tavares[5], R Thew[1], P Trinkler[2], S Ventura[6], R Voirol[6], N Walenta[1], H Zbinden[1]**

[1] University of Geneva, Group of Applied Physics, Rue de l'École de Médecine 24, CH-1205 Geneva, Switzerland
[2] ID Quantique SA, Rue de la Marbrerie 3, CH-1227 Carouge, Switzerland
[3] University of Applied Sciences Western Switzerland in Fribourg (EIA-FR), Boulevard de Pérolles 80, CH-1705 Fribourg, Switzerland
[4] ETH Zurich - Integrated Systems Laboratory, Gloriastrasse 35, CH-8092 Zurich, Switzerland
[5] University of Applied Sciences Western Switzerland in Geneva (hepia Geneva), Rue de la Prairie 4, CH-1202 Geneva, Switzerland
[6] University of Applied Sciences Western Switzerland in Yverdon-les-Bains (HEIG-VD), Route de Cheseaux 1, CH-1401 Yverdon, Switzerland

E-mail: `Damien.Stucki@unige.ch`, `Matthieu.Legre@idquantique.com`

**Abstract.** In this paper, we report on the performance of the SwissQuantum quantum key distribution (QKD) network. The network was installed in the Geneva metropolitan area and run for more than one and a half years, from the end of March 2009 to the beginning of January 2011. The main goal of this experiment was to test the reliability of the quantum layer over a long period of time in a production environment. A key management layer has been developed to manage the key between the three nodes of the network. This QKD-secure network was used by end-users through an application layer.

PACS numbers: 03.67.Dd, 03.67.Hk,

‡ D Stucki and M Legré contributed equivalently to the writing of this paper.

## 1. Introduction

Quantum random number generators [1, 2] have already been identified as the first technology resulting from quantum information science to reach the market and quantum key distribution (QKD) is closely following in its footsteps in this rapidly emerging field. Indeed, we have gone a long way since the first paper on QKD by Bennett and Brassard in 1984 [3] §. The reviews of Gisin et al. [5] and Scarani et al. [6] present the evolution of QKD over these last few decades. However, to definitively be a commercial success, QKD needs to demonstrate its *integration* in telecommunication networks, its *reliability* and its *robustness*.

For *integration*, QKD has to be adapted to the topologies developed in telecommunications networks for unicast, multicast and broadcast traffic. Unicast means point-to-point traffic. Multicast indicates traffic between a subgroup of nodes of the network. Broadcast denotes traffic shared between all the nodes of the network. As current QKD links are basically point-to-point links, the integration in telecommunications networks requires additional optical components and/or software. According to the requirements of the network and the present state of the art of quantum devices, two types of QKD networks can be implemented presently: either based on trusted intermediate nodes or additional optical components. Trusted-node networks allow one to expand the maximal distance of QKD, but require physically secure intermediate nodes. QKD networks based on optical components (active optical switches, circulators or passive dense wavelength division multiplexing, for instance) allow one to share infrastructure (fibre link, for instance) and do not need trusted intermediate nodes. However, with this type of QKD networks, the maximal distance and the bit rate are limited by the optical attenuation of the link. Note that the two types of nodes can be mixed. Thus, in the last few years, several QKD networks have been deployed and tested using trusted nodes and/or optical components [7, 8, 9, 10, 11, 12, 13, 14, 15]. These networks, with the exception of the network presented in [8], were deployed for short periods of time: at most a few months. The quantum layer of the SwissQuantum network relies on trusted intermediate nodes and ran for almost two years. Note that another kind of work has been done on the integration of QKD in optical fibre networks over the last years. This work focuses on the multiplexing of quantum channels with classical channels on a single fibre [16, 17, 18, 19, 20]. The main motivation for this is the reduction of the cost of QKD implementation by sharing a fibre for multiple applications and/or users, e.g. point to point QKD and classical communication with wavelength demultiplexing and encryption of Fibre To The Home (FTTH) communications over Gigabit Ethernet passive optical networks thanks to keys shared by QKD.

§ A paper of Wiesner on *quantum money*, written in the 70's but published only in 1983 [4], inspired Bennett and Brassard for their paper.

The most important prerequisite for the integration of QKD in a telecommunications network is *reliability* because networks run 24 hours a day, 7 days a week, and 365 days a year. Thus, new devices in networks - QKD for instance - must not degrade the quality of service. Untrusted telecommunications are preferred to no communication at all by the network community. So, to demonstrate the integration of QKD technology within communication networks, we need to show the *reliability* of this technology over long periods of time and in production environments.

As next prerequisite, QKD systems require *robustness* because they run no longer in a laboratory, but in a rough environment. People working in server rooms do not always handle QKD systems with the same care as physicists. For instance, the fibres are not handled as they should in the field environment, e.g. fibre ends are not always as clean as in laboratory conditions and this can impact on the losses strongly (for example, dirty connectors can introduce -3dB extra losses corresponding to more than 10 km of fibre). As written before, if the losses are too large, the secret key rate of QKD is strongly reduced and can even be reduced to zero. Classical communications don't suffer from such an effect because if the losses are too large, the signal can be regenerated thanks to Er-doped amplifiers for example. Thus, before installing a QKD system, the fibres have to be chosen very carefully to have low losses. Note that upon disassembling the network, it was found that the protective cover of one optical fibre was damaged. In spite of this, the system still ran correctly.

## 2. The SwissQuantum testbed

*2.1. Topology*

The topology of the SwissQuantum network is presented in figure 1. It consists of three nodes:

- Unige (University of Geneva),
- CERN (Centre Européen de Recherche Nucléaire),
- hepia (Haute École du Paysage, d'Ingénierie et d'Architecture),

and three point-to-point links:

- Unige - CERN,
- CERN - hepia,
- hepia - Unige.

Each node is divided in two sub-nodes, one for each point-to-point link connected to the node. The node at CERN is in France. The two other nodes are in Switzerland. Hence, the SwissQuantum network is the first international QKD network.

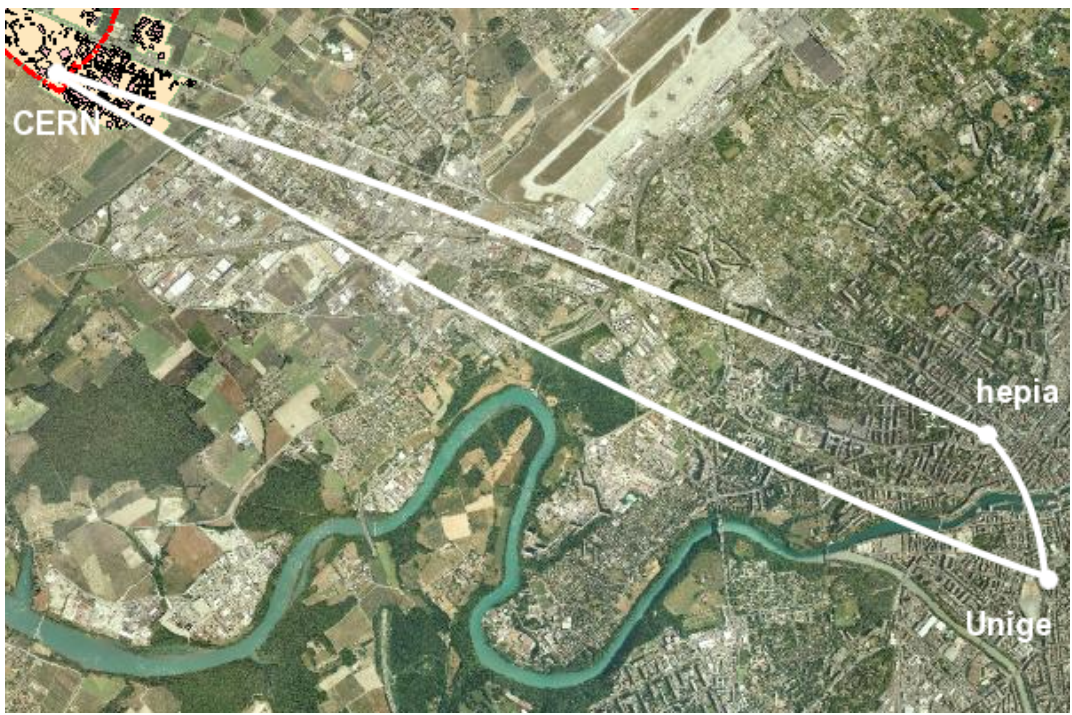

**Figure 1.** Map of the SwissQuantum network. Two nodes are in the Geneva city centre and the third one is on the site of CERN in France (the border is in red). The white lines are drawn for illustration: they do not represent the fibres.

*2.2. Structure*

The SECOQC network [10] introduces the idea of layers for QKD networks. The concept of layers allows one to add a mediation layer between the QKD layer and the secure application layer. This provides flexibility in the integration of QKD devices in telecommunications networks. For instance, in the SECOQC network, QKD servers implementing different protocols were associated thanks to the key management layer. The same type of configuration with three layers was implemented in the Tokyo QKD network [15].

The SwissQuantum network also consists of three layers (see figure 2):

- a quantum layer composed of QKD point-to-point links implemented with commercial QKD devices (ID Quantique, id5100) [21];
- a key management layer in charge of the management of secret keys across the network and between the layers;
- an application layer where the keys provided by the key management layer are used by the end-user applications.

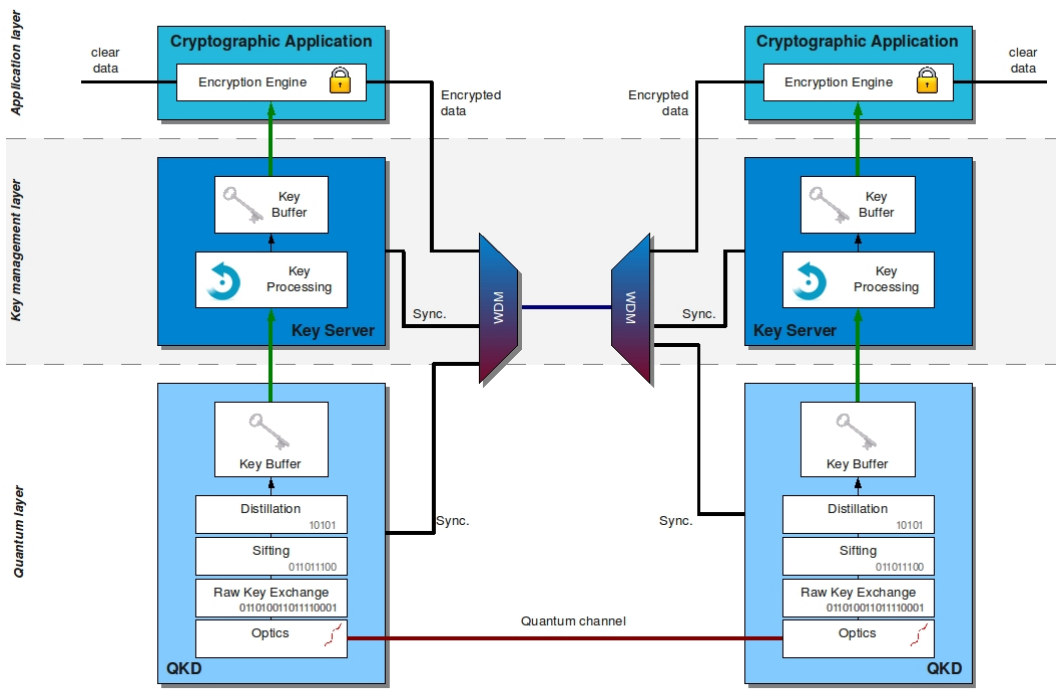

**Figure 2.** Structure of one point-to-point link with the quantum, key management and application layers. The quantum layer generates the secret keys, and pushes them to the management layer. The key management layer processes and stores the keys obtained from the quantum layer and pushes them to any applications which request secret keys (see text for details).

The different layers are presented in more detail in sections 2.4 to 2.6.

*2.3. Detailed layout of the SwissQuantum network*

Figure 3 shows the SwissQuantum network topology in more detail. There is one pair of dark fibres for each node connection, except for the connection between CERN and Unige. Between CERN and Unige, a pair of fibres is dedicated to the QKD link and one pair is dedicated to the data transmitted by the commercial 10 Gbps Ethernet encryptors (one fibre for each direction). The data transmitted by the 10 Gbps encryptors is real data, so it is separated from the QKD network to avoid data transmission interruption due to maintenance of the QKD network. Apart from the pair of fibres used by the 10 Gbps encryptors, one fibre of each pair of dark fibres is used as a quantum channel, whereas the other fibre is used for transmitting all the classical channels. Depending on the connection, the classical channels can be composed of the classical channel for the QKD system, the classical channel for the key servers, the classical channel for encryption applications, and/or the classical channel for the monitoring of all the devices. Each classical channel needs to work in both communication directions. In order to multiplex all the classical channels between two nodes in a single fibre, they are multiplexed using wavelength division multiplexing (WDM) techniques.

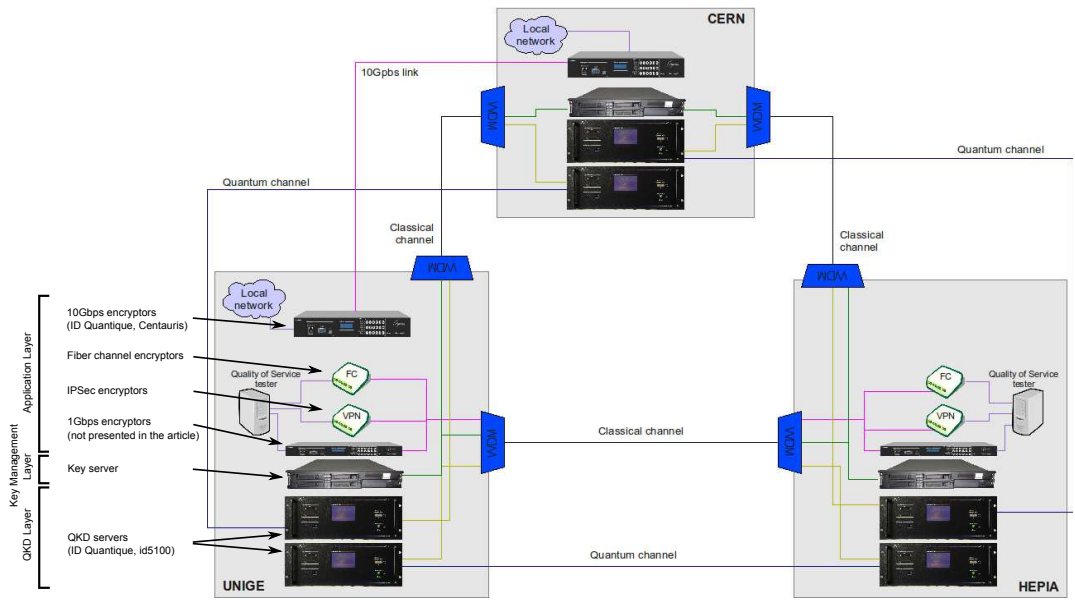

**Figure 3.** Detailed topology of the SwissQuantum network. All the lines are optical fibres for connecting the different apparatus.

One part of the network is missing in figure 3, namely the monitoring network.

Virtual local area networks (VLAN) were developed to monitor the SwissQuantum network. The three VLANs, one per layer, were connected to a server at hepia, which was in charge of the monitoring of the SwissQuantum network. Two firewalls were deployed: one to avoid illegitimate connections from the internet to the server; a second one to limit access to the management network. So, only legitimate entities (ID Quantique, Unige and hepia) could access to the monitoring network through ssh connections.

*2.4. Quantum layer*

The quantum layer consists of three point-to-point quantum links described in Table 1.

**Table 1.** Characteristics of the QKD links in the SwissQuantum network.

| Name | Nodes at the two ends | Length of fibre (km) | Optical loss (dB) |
|---|---|---|---|
| SQ1 | CERN - Unige | 14.4 | -4.6 |
| SQ2 | CERN - hepia | 17.1 | -5.3 |
| SQ3 | Unige - hepia | 3.7 | -2.5 |

Each quantum link is implemented with a pair of customized commercial QKD servers (ID Quantique, id5100 [21]). The optical platform of the QKD servers are based on the so-called plug & play configuration [22]. A simplified scheme of this go & return configuration is depicted in figure 4. The device on the left side of figure 4 consists of a Faraday mirror, a phase modulator and a variable optical attenuator. The other device is composed of an unbalanced Mach-Zehnder interferometer, two single-photon detectors and a laser preceded by an optical circulator. Note that the beamsplitter on the left side of the interferometer is a polarization beamsplitter. The main advantage of this optical platform is its intrinsic auto-compensation of phase and polarization fluctuations in the quantum channel. Indeed, the phase auto-compensation is guaranteed by the single interferometer which is used for the qubit preparation and analysis. The polarization auto-compensation is guaranteed by the combination of Alice's Faraday mirror with Bob's polarization beam splitter (more details can be found in [22]). In figure 2, this corresponds to the optics boxes, which generate the raw keys and push them to the sifting process. The raw key exchange process stops when the buffer registering the phases applied on Alice side is full or when the probability of detection on one or both of Bob's detectors is too low. For the links in the SwissQuantum network, this corresponds typically to 5 to 7 millions of detections for a full buffer. The sifted keys then follow the reconciliation process which generates secret keys.

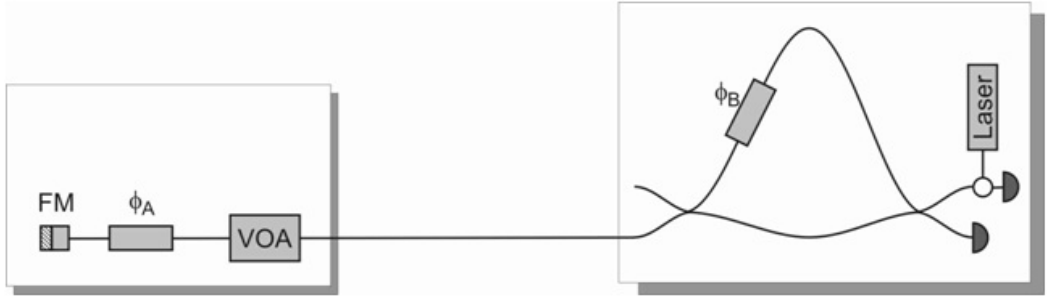

**Figure 4.** Simplified scheme of the plug & play QKD system. On the left (Alice): FM, Faraday mirror; $\Phi_A$, Alice's phase modulator; VOA, variable optical attenuator. On the right (Bob): the unbalanced Mach-zehnder interferometer with Bob's phase modulator $\Phi_B$; the laser the circulator and the two single-photon detectors.

The QKD servers can run with either standard BB84 [3] or SARG [23] protocols. The standard BB84 and SARG protocols differ only in the sifting part meaning that both protocols can be run on the same optical platform. This small difference allows SARG to be more robust against photon number splitting attacks. Hence, SARG is more efficient than standard BB84 over long distances. Within the SwissQuantum network, only the SARG protocol was used. The distillation of the secret keys is performed in three steps: error correction, privacy amplification and authentication of the classical communications. This distillation is performed each time Alice's buffer is full (5 to 7 millions detections, so 1.25 to 1.75 millions bits after sifting). The error correction is implemented using the Cascade algorithm [24]. The raw key buffer is split into blocks of 8192 bits that are corrected one after the other. In our implementation, there is no step of error estimation because we have the exact value of the quantum bit error rate (QBER) after the error correction. Cascade is robust enough to run efficiently even when it has only a rough estimate of the QBER value. The privacy amplification is done with the universal2 hash functions proposed by Krawczyk and based on Toeplitz matrices [25]. It is performed on the all-sifted buffer. The authentication is performed according to the Wegman-Carter scheme [26, 27]. All classical communications of a round of secret key exchange are authenticated at the same time. The raw key exchange and the distillation are done sequentially with typically 4-5 minutes for raw key exchange and 1 minute for the distillation. The quantum layer continuously generates secret keys and transfers them to the key management layer.

The initialization of the quantum layer is very important if we want to guarantee the security of the key exchange using QKD. Indeed, in order to avoid man-in-the-middle attacks, as written before, the classical communications need to be authenticated. In our implementation, this authentication requires secret keys to be performed. Those keys can be provided by the QKD if the quantum exchange has worked at least once. So, an initial secret needs to be shared by the two devices for the authentication of the first round of quantum key exchanges. Furthermore, in the case of an implementation with coherent weak pulses, the probability of detections must be monitored tightly to avoid photon number splitting attacks. Hence, the loss value of the quantum channel has to be known in order to first adjust the mean number of photon per pulse to its optimal value and then compute the expected probability of detection which is used as a reference for the measured detection probability monitoring. Both the initial secret key and quantum channel loss value, are stored in the QKD devices. These two parameters can be entered through the touch panel of the QKD servers (ID Quantique, id5100) (the blue screen in the middle of the front panel of the QKD servers in figure 3). This touch panel is protected by a standard authentication procedure using a password of at least 8 characters. Before the installation of the devices, they are set in a factory configuration which does not contain any loss value or initial secret. Those two parameters are asked during the installation procedure of the devices. Once the devices have been initialized, the quantum key exchange starts automatically and it is impossible to change the two parameter values without stopping the key exchange. To change these parameters, the system needs to be reset to the factory configuration.

Note that in this paper we do not consider quantum hacking. QKD has been proven information-theoretically secure. But, of course as any cryptographic technology, its security relies on security proofs and a correct implementation of the system. Quantum hacking exists for at least 10 years [28, 29] and has been particularly active for the last couple of years. Its importance is recognized by the research community since the middle of 2010 [30, 31]. The goal of quantum hacking is to show loopholes in given QKD implementations, and to propose countermeasures against those loopholes [32, 33, 34, 35]. The SwissQuantum project started before 2010, i.e. before that quantum hacking became an important aspect of QKD security. This is why did not include any patch against any of the attacks demonstrated since the launch of the project to avoid any interruption due to patching.

*2.5. Key management layer*

The key management layer is the interface between the quantum layer, where secret keys are generated, and the application layer, where secret keys are used. It is responsible for the processing of keys, their storage in each node, and their management between the nodes and the layers. It consists of one computer per node, called key server, with a buffer dedicated to key storage and a synchronization channel between each of them. This approach allows one to go from a very basic network topology composed of several point-to-point QKD links to more complex network topologies.

The SwissQuantum project focuses on network features linked to performance, flexibility and reliability. The main guideline we have followed is based on the quite recent concept of link aggregation. This concept is used to increase both the bandwidth and the availability of a link between two locations thanks to multiple network connections between these locations. Several standards on this method have been defined since 2000 [36], the latest one, appeared in 2008, is IEEE 802.1AX-2008 [37]. To explain operation of the link aggregation, let us consider a very simple configuration where two locations are connected through two optical cables. A switch in each location can direct the data traffic either in the first or in the second cable. Obviously, link aggregation allows one to increase the bandwidth by sending half of the traffic through the first fibre and the other half through the second fibre. If the receiving switch is able to recombine all the data together, the bandwidth of the link composed of two optical cables is two times higher than that of a single cable link. Furthermore, if one of the two cables is cut or unplugged, all the data can be redirected to the remaining active cable, providing greater network resilience. For this reason, we believe that quantum networks should have this kind of features. Thus, we have applied the link aggregation concept to the distribution of quantum keys. The main difference between QKD networks and classical networks is that in classical networks data is transmitted whereas in QKD networks secret keys are exchanged. It is extremely problematic to lose data, but a reduction of the key exchange rate has no impact on either data or security. This is why for QKD link aggregation we do not need active switches. The same buffers can be used on both sides to store the keys exchanged through the first and the second link. The applications do not need to know if the keys they are using have been exchanged through link one or two, but they need to get the same keys on both sides. If one of the two links is down, there are still keys exchanged through the other link. The rate of secret keys stored in the buffers is the sum of the rate of the keys exchanged through the first link and the rate of the ones exchanged through the second link. Our implementation of QKD link aggregation doesn't require active switches, but as many QKD systems as the number of links between the two locations. Indeed, our QKD link aggregation implementation needs two sets of QKD devices, one for each link. More technical information on the implementation can be found in section 3.

In addition to the link aggregation configuration, parallel key agreement was implemented. The parallel key agreement consists of the combination of secret keys obtained by independent processes. In the key management layer of the SwissQuantum network, the simplest version of parallel key agreement was implemented: dual-key agreement. The keys exchanged with quantum cryptography and keys exchanged with help of a Public Key Infrastructure (PKI, [38]) are combined to obtain the final key. Depending on the combination technique, this final key can be as secure as the more secure of the two initial keys. PKI relies on asymmetric key cryptography. The security of asymmetric cryptography has not been proven according to information theory. This combination is not used to increase the security of the resulting key, but for improving the reliability and availability of the applications in case of failure of the QKD layer: if users can accept data transmission with a security limited by the conventional key exchange technique, they can avoid stopping all applications. This method can be seen as a way to improve the availability of the link. Moreover, dual-key agreements allow the use of keys generated by QKD devices to be certified, which is required for some applications. Unfortunately, the certification for QKD devices themselves does not exist yet, but should be available in the near future thanks to the work on standards for QKD by the Industry Specification Group initiated by the European Telecommunications Standards Institute [39].

In summary, the SwissQuantum key management layer implements the following functionalities :

- key redundancy - multiple paths for key generation;
- increase of key generation speed - use of multiple QKD paths for providing keys to the same application link;
- dual-key agreements - combination of the keys obtained by two independent

key agreement techniques to generate a resulting key;
- key storage - secure buffers to store the keys.

A detailed description of the implementation is given in section 3.

## 2.6. Application layer

The application layer is the layer where the keys produced by the quantum layer and handled by the key management layer are employed by the final user. It consists of the connection of conventional network devices like switches, routers or encryptors. This application layer is independent from the quantum and the key management layers, except for the key requests. All applications requiring secret keys can make a request to the key server located in the same node. The reliability and availability of this layer is very important, this is the reason why the dual-key agreement, as described in the previous section, was implemented in the key management layer. Dual-key agreement allows one to run the application layer continuously, even if the quantum layer can not generate any key for some short period of time. Within the SwissQuantum network, we implemented several QKD-enhanced encryption applications at both layers 2 and 3:

- 10 Gbps Ethernet encryptors (Layer 2);
- 2 Gbps Fibre Channel device encryptors (Layer 2);
- IPsec encryptors (Layer 3).

Layers 2 and 3 refer to the standard network layers as defined by the Open Systems Interconnection (OSI). Layer 2 is the data link layer, the layer carrying the Ethernet frames, for instance. Layer 3 is the network layer which carries the IP packets. The main advantages of performing the encryption in Layer 2 are that firstly, the encryption does not reduce the bandwidth, and secondly, the latency introduced by the encryptors is very small. Doing the encryption on Layer 3 strongly reduces the bandwidth of the link because of the need of encapsulation (addition of extra header and footer to the frame). Furthermore, in general, a large latency is introduced by Layer 3 encryption due to its implementation which is done with a microprocessor. However, Layer 3 encryption is more suitable when the traffic goes through network components that work on Layer 3. Moreover, Layer 3 encryption (software implementation) is less expensive than Layer 2 encryption (hardware implementation). Hence, each layer has advantages and disadvantages. The SwissQuantum network demonstrates the versatility of QKD by the integration of both Layer 2 and Layer 3 devices.

*2.6.1. 10 Gbps Ethernet encryptors (Layer 2)* Commercial high-speed Layer 2 encryptors (IDQ, Centauris [40]) that are compatible with QKD performed 10 Gbps Ethernet encryption. They implement the Advanced Encryption Standard (AES, [41]) using a key size of 256 bits and support multiple protocols, among them Ethernet up to 10 Gbps. These encryptors work with the dual key agreement between an internal key exchanged via PKI and an external key. The dual key agreement is done in such a way that the encryptor using it is FIPS 140-2 level3 certified [42]. The exchange of session keys via PKI between two encryptors is achieved by means of X.509 certificates. Certificates are a form of electronic credential (like a passport) endorsed by a trusted third party certifying authority (CA). Each certificate contains an identifying name, unique serial number, expiry date and public key and prior to installation is signed by the CA.

*2.6.2. 2 Gbps Fibre Channel encryptors (Layer 2)* The QKD-enhanced encryption and authentication device [43, 44] performs high-speed 2 Gbps encryption and authentication of the data at Layer 2. The encrypted and authenticated data are sent using the Fibre Channel transport mode. These encryptors support the dual-key agreement, which has been implemented in a similar way than for the commercial 10 Gbps Ethernet encryptors.

*2.6.3. IPsec encryptors (Layer 3)* The QKD-enhanced IPsec encryptor integrates the cryptographic symmetric key generated using the QKD protocol with the IPsec suite of protocols, in order to provide a point-to-point, quantum-secure communication link operating at Layer 3.

## 3. Details on the implementation of the key management layer

*3.1. Requirements on the network*

The CERN - Unige link was privileged in the SwissQuantum network. Thus, the architecture and implementation of the SwissQuantum network was developed such as to reduce as much as possible the risk of losing the availability of this link. To ensure this, firstly, we used commercial devices to perform the encryption on this link, secondly, we implemented the key management layer in such a way that the CERN - Unige link was favoured in relation to the two other links.

*3.2. Implementation of the key management layer*

A scheme of the implementation of the key management layer is depicted in figure 5. The three nodes are implemented in different manners. The three quantum key exchange links are represented by dashed black lines. Two of the three connection links carried encrypted data: CERN - Unige and Unige - hepia (thin dark line in figure 5). The commercial 10 Gbps Ethernet encryptors were installed between CERN and Unige. The 2 Gbps Fibre Channel encryptors and IPsec encryptors were tested between Unige and hepia. There is one key server in each node, which manages the storage and distribution of the secret keys in several key buffers. Each key buffer is dedicated to a single application.

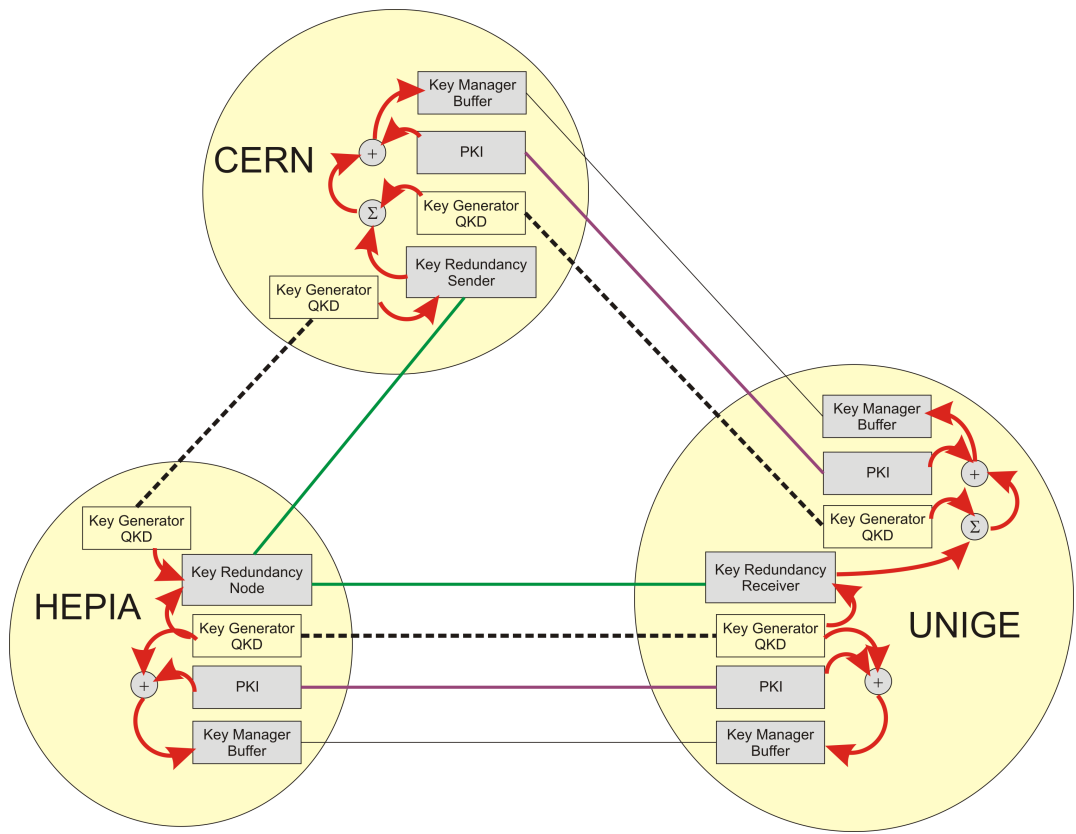

**Figure 5.** Key management layer implementation. All the functions in grey boxes whithin the same node are implemented in the same Key Server, or PC.

As explained above, the CERN - Unige link was privileged and hence we used the QKD link aggregation scheme to assure redundancy of the key exchange between CERN and Unige. This means that the key buffers of the key managers on this link were filled up with secret keys exchanged through the direct QKD link, and with secret keys distributed through the intermediate node (hepia) over the two other QKD links (CERN - hepia and hepia - Unige). The key distribution through the intermediate node is represented by the two green links in figure 5. A key redundancy sender in CERN generates a random key K, which is encrypted with the One-Time-Pad (OTP) protocol. The key used for the OTP encryption is one of the keys exchanged by the QKD devices between CERN and hepia. The encrypted key K is sent to the key redundancy node in hepia. This encrypted key K is decrypted by the key redundancy node and then encrypted with a key shared between hepia and Unige by QKD. It is sent to the key redundancy receiver in Unige, and decrypted. At the end of this process, the key K has been exchanged between CERN and Unige through the intermediate node hepia. The keys exchanged through the intermediate node are concatenated with the keys exchanged through the direct link. This concatenation of the keys is represented by the sign '$\sum$' in figure 5.

Before storing the secret keys in a buffer, an internal dual-key agreement is performed. The implementation of this PKI is similar to the one in the commercial 10 Gbps encryptors, hence it follows the recommendations of the X.509 standard and is based on RSA cryptographic scheme. As previously stated, the PKI allows one to maximize the availability of keys for the secure applications. Some keys are exchanged between the pairs of key managers using PKI (purple links in figure 5). These keys are combined with keys provided by the QKD devices using an XOR operation. This operation is represented by a '+' sign in figure 5. The resulting key can be seen as the cyphertext of the PKI key encrypted using OTP with the QKD key. OTP is proven information-theoretically secure if it is used with perfectly random secret keys and each key is used only once. The keys exchanged through QKD have been proven information-theoretically secure. This means that the mutual information between the PKI key and the resulting key is equal to zero. Hence, knowing the PKI key does not give any information on the resulting key. Moreover, since the QKD key is random and independent on the PKI key, the resulting key is random too. So, even if the PKI key is entirely known to the adversary, the resulting key is secure. The resulting keys are stored in the key buffers and are sent by the key managers to the applications each time a new key is needed.

## 4. Long term performance of the quantum layer

In telecommunication networks, one of the most important figures of merit is the bit rate. By analogy, the secret key rate is considered as the key parameter for QKD devices. The secret key rate is derived from the raw key rate and the quantum bit error rate (QBER). Thus, the probability of detection - giving the raw detection rate by multiplying it by the number of gates per second - and the QBER measurement are reported, as well as the secret key rate, in this section.

*4.1. Probability of detection*

Figure 6 presents the probability of detection for the single-photon detector 1 of the different systems. The probability of detection is the probability of having a detector click - due to a photon or a dark count - per gate of activation of the detector. The probability of detection is calculated over the time period needed to fill the buffer with raw detection data.

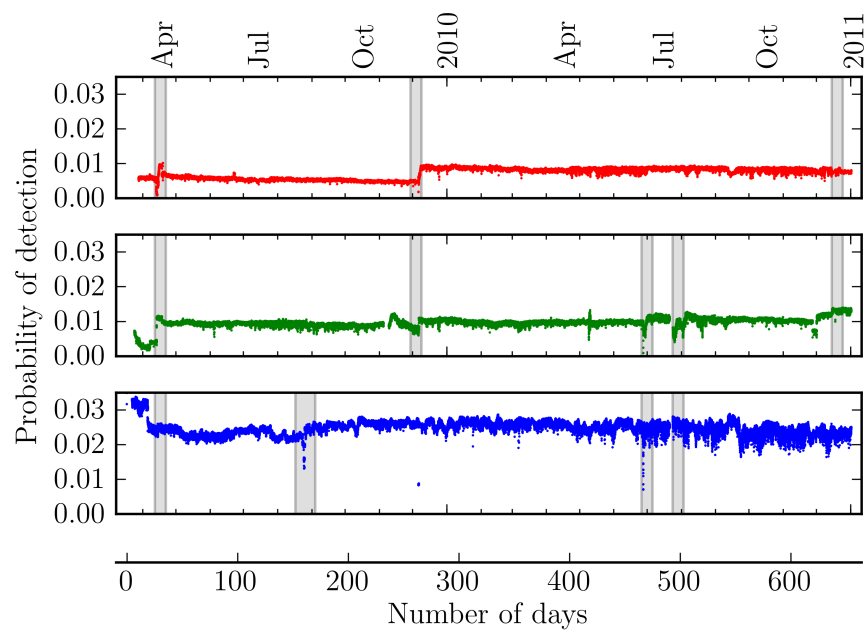

**Figure 6.** Probability of detection as a function of time for detector 1 of SQ1 (top), SQ2 (middle) and SQ3 (bottom) links. Detector 2 has similar behaviour. (The grey areas are guides to find the perturbations described in the text.)

Although the probability of detection was rather stable over the 21 months, there were several perturbations or variations which can be seen in figure 6. We assign those variations to two different classes. The first class corresponds to a long-term change of the mean detection probability. The second class of perturbations corresponds to short-time changes.

The long-term variations of mean detection probability value levels can be explained as follows:

- after about 15 days, the probabilities of detection of the different systems was optimized to maximize the secret key rate after initial test phase (SQ1 to SQ3);
- after a power cut at CERN around day 260, we took the opportunity to change some settings in the single-photon detectors of the three systems. Indeed, we noticed during the first 260 days of working that the temperature in the IT rooms at hepia and Unige was not always below 30°C. ID Quantique systems are specified to work at a maximal room temperature of 30°C. This limit is due to the cooling capability of the APDs in the single-photon detection modules. To reduce the risk of having the system stopped due to a too high room temperature, we have changed the cooling temperature from -50°C to -40°C. After changing the temperature, we tuned the detector efficiency at values close to the previous ones but slightly different. This explains the differences in the mean detection probability values measured before and after this intervention especially for SQ1 and SQ2.

The short-term interruptions/reductions of detection probability are mainly due to external problems and not directly to the quantum layer. The most important problems encountered are listed below:

- 20th of August 2009 to 2nd of September: bug in the software handling the communication between the QKD servers and the key servers. The bug appears only on the SQ3 link. It was fixed for all systems.
- 2nd of December 2009: power cut at CERN, SQ1 and SQ2 links down for about 8 hours.
- 29th of June 2010: air conditioning problem at hepia: it was not possible to maintain the temperature of the single-photon detectors in Bob (SQ2 link) at -40ºC with an external temperature of 45ºC. SQ2 link down for few hours. There was no problem with Alice (SQ3 link) except a small reduction of the bit rate due to a small decrease of the visibility.

- 27th of July 2010: general maintenance problem at hepia leaving all systems in the server room without power for the weekend. SQ2 and SQ3 links were down for the weekend. However, it took a few days to recover the stability because of the maintenance activities in the server room (not on the SwissQuantum systems).
- 18th of December 2010: power cut at the CERN during a weekend. SQ1 and SQ2 links down for the weekend.

*4.2. Quantum bit error rate*

The QBER was also recorded during the full period of the experiment (see figure 7). The fluctuations observed are due to statistical fluctuations and detection probability fluctuations. Nevertheless, the value of the QBER was pretty low and stable during the 21 months.

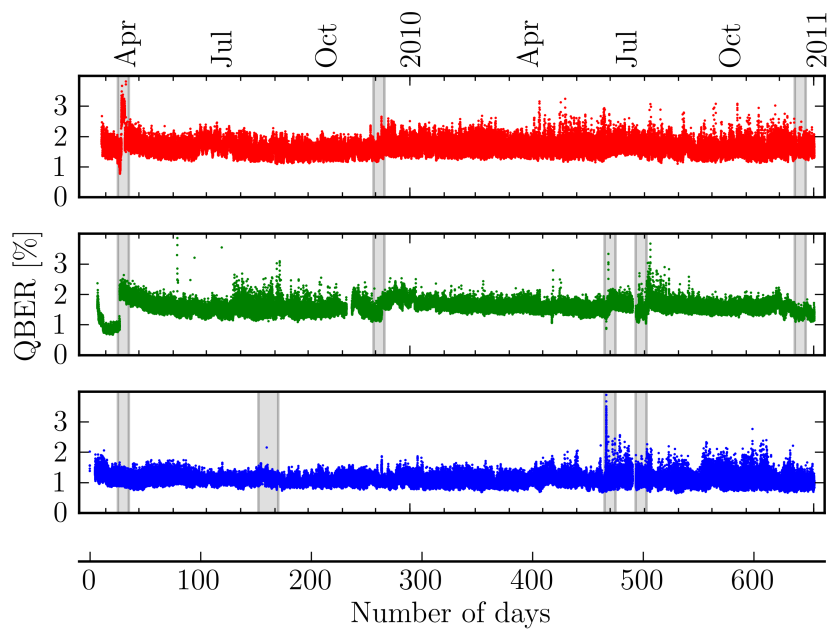

**Figure 7.** QBER as a function of time for SQ1 (top), SQ2 (middle) and SQ3 (bottom) links.

*4.3. Secret key rate*

For the final user of the application layer, the most important parameter is the number of keys that he can use for its applications. In the SwissQuantum network, the devices employed 256-bit keys. Figure 8 presents the number of 256-bit keys generated per day. This rate is quite stable over more than 600 days. Thus, the QKD systems deployed in the SwissQuantum network proved the robustness of QKD for long-term deployment in telecommunications networks. As explained above, some interruptions in the secret bit rate generation have been observed. They are mainly due to external reasons, and the systems always recovered when the environment conditions went back to normal.

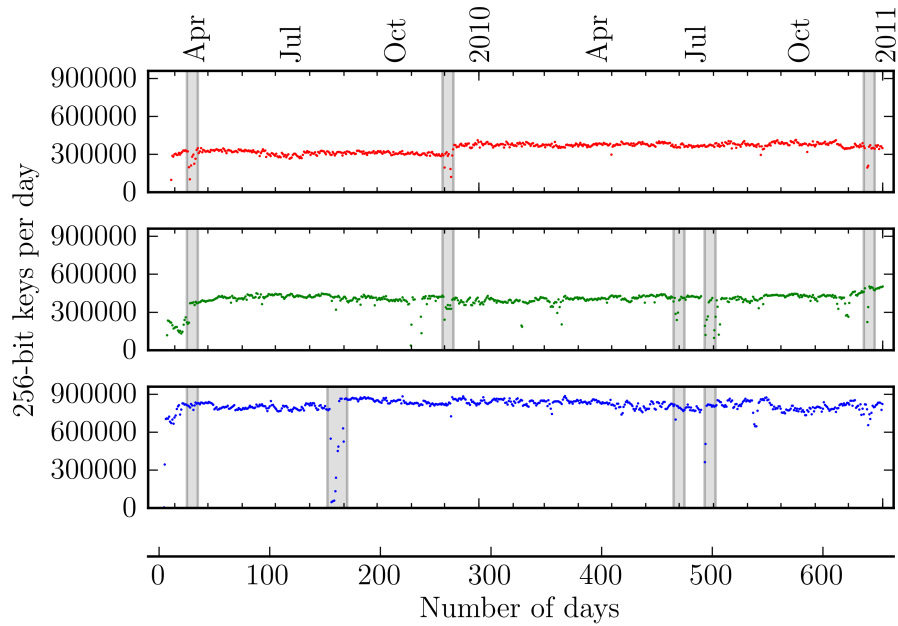

**Figure 8.** Number of 256-bit keys per day. The number of 256-bit keys created per day is larger for the SQ3 link, because the optical attenuation of this link is lower than for the SQ1 and SQ2 links at similar detector efficiencies. The number of keys per day for the SQ1 and SQ2 links are similar.

*4.4. Variation of the optical fibre length*

For the stability of the QKD layer over the 21 month period, the adaptation to the variations of the optical path is crucial. The length of the optical path is important for most QKD implementation schemes and especially for the plug & play one, because it defines when the single-photon detectors have to be activated to detect the photons coming from Alice [22]. The optical path changes due to variations of the physical length and/or the refraction index of the optical fibre. These changes are essentially consequences of temperature variations. Figure 9 presents the variation of the optical path length (go&return) and the temperature in Geneva. The absolute optical length variation is calculated relative to the mean optical length over the 21 months.

As can be seen in figure 9, the variations of the optical length and the temperature are very similar. The absolute deviations are larger for the SQ1 and SQ2 links than for the SQ3 link, since the fibres of SQ1 and SQ2 links are longer than the one of the SQ3 link and the variations of the optical length are proportional to the fibre length. In sub-figure (a), the seasonal variations are presented. The amplitude of the variations is 6 meters corresponding to 30 ns in optical fibre. As the width of detection gates is shorter than 2 ns and the width of the laser pulse is shorter than 1 ns, the length of the optical path has to be monitored to adjust the activation time of the detectors [22]. Sub-figure (b) gives the variations over three typical days. There is a shift of few hours between the optical length variations and the temperature. This is due to the inertia of the system. The temperature is measured in air, but most of the fibres are under ground. The important fact to underline is that the QKD devices are sufficiently flexible to automatically follow the variations of the optical path length.

## 5. Performance of the application layer

The test of the performance of application layer was not the primary goal of the SwissQuantum network, however, some tests were done. On the link CERN-Unige, the commercial encryptors (ID Quantique, Centauris) worked perfectly for the full duration of the SwissQuantum network with transmission of real data on the link. The 2 Gbps Fibre Channel encryptors (Layer 2) and the IPsec encryptors (Layer 3) were tested with Fibre Channel and Ethernet Test Modules (Exfo, FTB-8525/8535 Packet Blazer) racked in Universal Test System (Exfo, FTB-400). The systems were tested over months and the performances where in the specifications of the

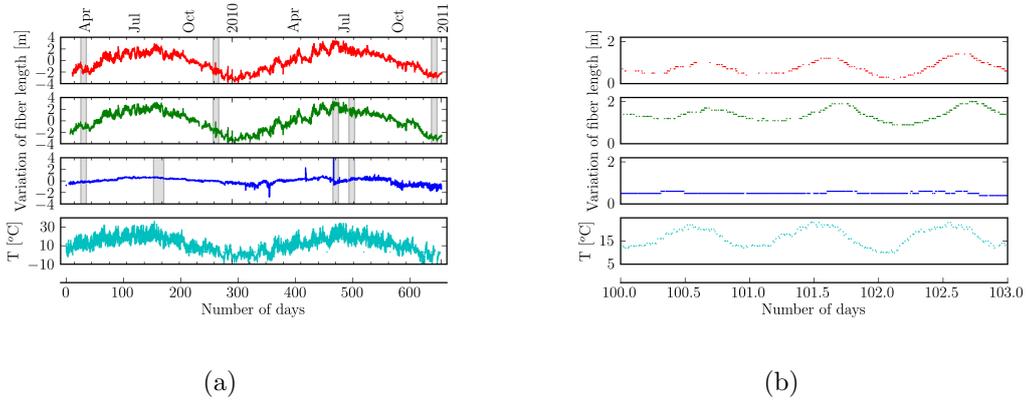

**Figure 9.** Variations of the optical fibre length and temperature in Geneva for go&return path. (a) 21-months measurement. (b) 3 days zoom

2 Gbps Fibre Channel and the IPsec protocols. The 256-bit key were changed each minute for the different encryptors. The change of key in the commercial encryptors was done without loss of bandwidth. The 2 Gbps Fibre Channel required 100 ns to change the key. For the IPsec encryptors, no measurement was done on the latency introduced by the key change because it has a negligible effect compare to the intrinsic throughput reduction due to encapsulation. For example, we measured an output throughput of IPsec varying from 10 to 95% of the input throughput depending on the frame size.

## 6. Conclusion

The SwissQuantum network demonstrates that QKD has the maturity to be deployed in telecommunications networks. It has proven its reliability and robustness in a real life environment outside of the laboratory. Furthermore, it shows that QKD technology can be integrated in quite complex network infrastructures. Those networks need a layer which makes the interface between the QKD layer (layer where the secret keys are exchanged) and the application layer (layer where the keys are used by the secure applications). Within the SwissQuantum project both the QKD layer and the interface layer, called key management layer, have run for more than one and a half year. The key management layer implemented in the SwissQuantum network takes over the concept of link aggregation. It allows the increase of the bandwidth and availability of secret keys between two locations connected through several links.

**Acknowledgments**

Financial support is acknowledged from the "NCCR Quantum Photonics" project, the "Hasler Foundation", the "Banque Privée Edmond de Rothschild", "armasuisse", the "Swiss National Science Fund", the "CTI 8483.1 NMPP-NM" project, the "Nanoterra Qcrypt" project, the FP7 European projects "Q-Essence" and "QuReP". The authors thank: David Crisinel, Pierre Durand and Gérald Ineichen of CTI Genève, and Edoardo Martelli of CERN for their help for the deployment of the SwissQuantum network, the CTI Genève for access to their fibre links and Greg Schinn of EXFO for the loan of Fibre Channel and Ethernet Test Modules (Exfo, FTB-8525/8535 Packet Blazer).